# TOPOLOGICALLY PROTECTED SUPERCONDUCTING RATCHET EFFECT GENERATED BY SPIN-ICE NANOMAGNETS.

V. Rollano[1], A. Muñoz-Noval[2], A. Gomez[3], F. Valdes - Bango[4,5], J. I. Martin[4,5], M. Velez[4,5], M. R. Osorio[1], D. Granados[1], E. M. Gonzalez[1,2], and J. L. Vicent[1,2].

[1] IMDEA-Nanociencia, Cantoblanco, 28049 Madrid, Spain.

[2] Departamento Física de Materiales, Facultad CC. Físicas, Universidad Complutense, 28040 Madrid, Spain.

[3] Centro de Astrobiología (CSIC-INTA), Torrejón de Ardoz, 28850 Madrid, Spain.

[4] Departamento de Física, Universidad de Oviedo, 33007 Oviedo, Spain.

[5] CINN (Universidad de Oviedo-CSIC), 33940 El Entrego, Spain.

Abstract

We have designed, fabricated and tested a robust superconducting ratchet device based on topologically frustrated spin-ice nanomagnets. The device is made of a magnetic Co honeycomb array embedded in a superconducting Nb film. This device is based on three simple mechanisms: i) the topology of the Co honeycomb array frustrates in-plane magnetic configurations in the array yielding a distribution of magnetic charges which can be ordered or disordered with in-plane magnetic fields, following spin-ice rules; ii) the local vertex magnetization, which consists of a magnetic half vortex with two charged magnetic Néel walls; iii) the interaction between superconducting vortices and the asymmetric potentials provided by the Néel walls. The combination of these elements leads to a superconducting ratchet effect. Thus, superconducting vortices driven by alternating forces and moving on magnetic half vortices generate a unidirectional net vortex flow. This ratchet effect is independent of the distribution of magnetic charges in the array.

**INTRODUCTION**

Ratchet effect names the unidirectional motion of out-of-equilibrium particles when they move on a landscape with asymmetric potentials. This net flow of particles does not need of being driven by applied forces with non-zero average strength. Ratchet effects are in the core of distinct scenarios, for example in the biological mechanism by which proteins are transported (protein translocation) to the appropriate destinations (1, 2) or in the transport of colloid particles (3, 4). Up to date, different types of ratchets have been studied (5-10). It is worth noting that ratchet mechanisms are based on periodic asymmetric barriers or wells which could be, at first sight, an impediment to "particle" motion, but conversely these obstacles are crucial to yield particle net motion.

Nowadays, nanotechnology provides the tools to mimic, in some way, ratchets found in nature. Ratchet effect has been proved in the framework of cooperative phenomena as magnetism (11-16) and superconductivity (17-21). Two basic ingredients are needed to obtain a ratchet device: 1) Input signals yielding fluctuating motion of particles with zero-average oscillations; 2) Periodic structures which lack of reflection symmetry. Superconducting vortices are a good choice to investigate ratchet phenomenology of interacting particles. If vortices are driven by alternating forces the first ingredient is fulfilled. Regarding asymmetric potentials, two different approaches have been studied: i) geometric periodic potentials (18, 19, 21 - 24); ii) magnetic periodic potentials (25-27). The former produces robust ratchets, but the asymmetric potentials cannot be manipulated. Conversely, magnetic induced potentials could be manipulated, but, at the same time, the ratchet performance could be jeopardized by outside factors as, for instance, demagnetization effects or applied magnetic fields.

In this work, we have design a robust and resilient ratchet device, based on non-periodic and asymmetric magnetic potentials, which can be changed without losing its ratchet function. The key factor is the use of topologically protected asymmetric magnetic potentials (to provide a robust ratchet effect) arranged within a spin-ice system (to provide configuration flexibility). We have to point out that spin-ice magnets (28) have arisen as a convenient and powerful tool to explore many interesting and exotic fields as magnetic monopoles (29). Superconducting vortices have been employed also to mimic spin-ice configurations (30-32) and very recently a reprogrammable flux quanta diode has been realized using vortices and spin-ice magnets (33). In our study, we have used spin-ice magnets in honeycomb geometry and superconducting vortices to obtain a robust and flexible ratchet. More interesting, the asymmetric potential origin is not the well-known asymmetric magnetic potentials connected to magnetic dipoles (25-27, 33); in our case, a new ratchet mechanism emerges related to a specific topological defect characteristic of patterned magnetic nanostructures (34, 35): magnetic half vortices composed of a pair of charged Néel walls. These half vortices are confined to the sample edge in the holes of the honeycomb lattice retaining their asymmetric character even in disordered configurations, and therefore, protecting the ratchet effect. The paper is organized as follows: In the next section the fabrication of the sample, micromagnetic simulations, magnetic force microscope (MFM) and experimental techniques are described. Then, we discuss the relevant facts related to our specific spin-ice topologically protected system. After that, we show that the superconducting vortex dynamics can be controlled. Then, the experimental ratchet effect data are shown and analyzed. Finally, a summary closes the paper.

**EXPERIMENTAL METHODS**

The cobalt (Co) based spin-ice geometry is fabricated by a combination of electron beam lithography and magnetron sputtering on a Si substrate. The honeycomb array is made of stripes of sputtered Co film with side length 300 nm, width 150 nm and thickness 20 nm. These dimensions, shown in Figure 1(a), have been chosen to ease the superconducting vortex control. After lift-off, a 100 nm thick Niobium film is sputtered on top of the array. By means of photolithography and reactive ion etching, the device is patterned into a cross-shaped bridge to allow magnetotransport measurements. More details regarding the fabrication process can be seen in (36).

Magnetic configurations at remanence were obtained from micromagnetic simulations performed with the finite difference code MuMax$^3$ (37) in order to compare with experimental Magnetic Force Microscope (MFM) images. The unit cell of the honeycomb Co lattice was discretized into cells of dimensions 4×4×2.5 nm$^3$ and repeated using periodic boundary conditions to generate the honeycomb lattice. Typical material parameters have been used for Co: $M_s$=1.4 ×10$^6$ A/m, A=3×10$^{-11}$ J/m, and K=0 J/m$^3$, being $M_s$ the saturation magnetization, A the exchange constant, and K the in-plane anisotropy. Polycrystalline cobalt presents a low in plane anisotropy K = 10$^4$ J/m$^3$, much smaller than shape anisotropy of the nanostructures, so that it is usually neglected in micromagnetic simulations (38). MuView code was used for visualization (39). MFM contrast was simulated from the calculated micromagnetic configuration at 50 nm lift height. Domain structure was characterized by Magnetic Force Microscopy (MFM) at remanence with a Nanotech™ Atomic Force Microscope system with magnetic Nanosensors™ PPP-MFMR commercial cantilevers (spring constant 3 N/m). Measurements were performed in dynamical retrace mode at constant lift height (30 - 50 nm) over the topography profile acquired previously (40). Magnetotransport measurements were carried out on a commercial He cryostat with a superconducting solenoid (with magnetic fields up to 9 T). The sample is mounted in a computer controlled rotatable sample holder that allows applying in plane magnetic fields to the sample (modifying the magnetic history of the hybrid sample) or perpendicularly to the sample plane (tuning the density of superconducting vortices in the sample). Magnetotransport measurements are carried out with the input currents applied in the direction perpendicular to one of the easy axes. Therefore, the vortex motion is parallel to easy axis. The electrical characterization was performed applying an (ac) alternating (1 kHz frequency) or direct (dc) input currents and measuring the output dc voltages using commercial instrumentation; for more experimental details see (36).

**EXPERIMENTAL RESULTS**

a) **Magnetic characterization**

In a recent publication, Loehr et al. (41) show that the motion of colloidal particles can be controlled in a substrate with a hexagonal garnet film. The result is a topological protection which turns out in a robust transport of the colloidal particles which remains unchanged versus modification in sizes, mobility and magnetic susceptibility. Following this trend, we have chosen a Co honeycomb structure (see Fig. 1). In this section the magnetic characterization of our Co honeycomb nanostructure is present. The connected Co bars obey a particular case of usual spin-ice rules (42-43), thus, in our honeycomb sample the magnetization directions follow the so-called pseudo spin-ice rules (44-49): Two in – one out or one in – two out. We will see that in our device the combination of these two features, topology and spin-ice, is crucial to obtain a topologically protected vortex ratchet effect.

We begin describing the particular magnetism of the honeycomb array (see Figure 1(b)), focusing on the distinctive magnetic state in the vertices of the array. A simple and ordered magnetic configuration can be obtained at remanence when the saturating magnetic field is applied along one of the three magnetic easy axes of the structure; that is, parallel to any to the three nanobar directions of the honeycomb pattern. This can be seen, for example, in the micromagnetic simulation of Figure 1(b) for a field applied along the vertical axis of the array. In our case the applied saturating magnetic field was 7 T. In the remanent magnetic state the magnetization lies parallel to each of the bars in the image, surrounding the hexagonal holes of the honeycomb pattern, so that the remanent magnetization $M_R$ is parallel to the saturating field direction $H_S$. This magnetic configuration can be described with two distinct but related topological descriptions depending on whether we focus on the dipolar orientation of each bar in the array (spin ice charges (42-43)) or we focus on the detailed micromagnetic configuration at each vertex (Néel walls and magnetic half vortices (34-35)).

Starting with the former: the dipolar description (this is best observed in the simulated MFM image of Fig. 1(c) and in the experimental MFM image of Fig. 2(a)), we observe white or black contrast regions at each intersection of the honeycomb lattice arranged in two interleaving triangular lattices. The different magnetic contrast is created by the net magnetization divergence in each kind of intersection: a) white regions correspond to magnetization pointing into the intersection at one of the bars and out in the other two (see sketch in Fig. 1(c)), that is, to one-in/two-out (-1 spin-ice charge); b) black regions correspond to magnetization pointing into the intersection at two of the bars and out in the remaining one, that is to a two-in/one-out (+1 spin-ice charge). The ordered arrangement of black/white spots (+1/-1 spin-ice charges) found in Figs. 1 and 2(a) is the Ice II type (44-49).

Next, if we turn our attention to the local micromagnetic configuration, we observe that Néel walls are generated at the intersections between bars to accommodate the 60º magnetization rotation needed to follow the direction imposed by bar geometry. Magnetic half vortices are found at the points in which a V-shaped pair of Neel walls meets at the sample edge. There is one at each side of the vertical bar with magnetization aligned with $H_S$ (and parallel to $M_R$). At both magnetic half vortices there is a $-\pi$ (counter-clockwise) magnetization rotation corresponding to -1/2 topological index (34, 50). These magnetic half vortices correspond to black/white regions observed both in the experimental and simulated magnetic force microscopy (MFM) images (Fig. 1(c) and Fig.2(b)). The divergence of the magnetization associated to the magnetization rotation at the charged Néel walls generates the stray fields that will provide a magnetic potential for superconducting vortices. Fig. 1(d) shows the simulated contrast profile upon crossing a vertical bar bounded by two half vortices from bottom to top of the image (i.e. along the direction defined by remanent magnetization). The profile shows an attractive well (between points A and B), corresponding to the black half vortex, and a repulsive hill (between points C and D), corresponding to the white half vortex. Taking into account that pinning forces are given by potential gradients we observe that the asymmetry in the potential is the same in both cases: if forward direction is defined from A to D (i.e. by the remanent magnetization direction) the gradual ascending slopes (A'B and CD') correspond to small backward pinning forces whereas the steep descending slopes (AA' and D'D) correspond to large forward pinning forces. A'B and CD' can be associated to the broad tails of the Neel walls and AA' and D'D to the narrow cores. Then, the intrinsic asymmetry of the magnetic potential can be estimated from the width of the Neel core $W_{core} = 2(2A/\mu_0 M_s^2)^{1/2}$ in comparison to the width of the Neel tail $W_{Tail} = 0.56t\ (\mu_0 M_s^2/2K)$ (51), which for a Co film of thickness t = 20 nm, is of the order of $W_{core}/W_{tail}$ = 10 nm/1 μm = 0.01. The simulated profile shows a reduced asymmetry $W_{core}/W_{tail}$ = 0.25 due to a broadening of the effective domain wall core by convolution with the stray field from the MFM tip and to the confinement of the domain wall tails by the patterned honeycomb structure. In any case, we arrive at two important conclusions: first, the asymmetric potentials are linked to each of the individual half vortices in the bar; i.e. they do not depend on any specific sequence of +1 and -1 charges. This is; the asymmetry origin is not related to magnetic dipole as reported before (25-27, 33). Second, the sign of the asymmetry is the same for the black and white half vortices, and it is correlated in the whole honeycomb array by the magnetization rotation, clockwise or counter-clockwise, imposed by array geometry around the hexagonal holes. Therefore, the specific topology of the array is the clue for reaching this magnetic configuration.

In conclusion, combining these two approaches (micromagnetic and spin ice), we can describe the magnetic configuration of the Co honeycomb lattice in terms of two kinds of -1/2 magnetic half vortices, either associated with a +1 ice charge (black half vortex) or with a -1 ice charge (white half vortex); and interestingly each vertex contains two charged Néel walls.

Finally, spin ice geometry will allow us to study what happens when we disorder the magnetic potentials. Disorder can be easily introduced in the honeycomb Co lattice by changing the magnetic history with a variety of possible metastable configurations. Ice I states, for example, are characterized by a random mix of -1 and +1 spin ice charges (i.e. of negative/positive magnetic charges at the intersections of the honeycomb lattice). For example, if we apply a 7 T saturating magnetic field in the hard direction, i. e. perpendicular to one of the bar directions, the MFM image reveals a disordered remanent magnetic state, as shown in Fig. 2(c), in which black and white magnetic charges are randomly intermixed. The intensity of the MFM signal is very similar in all the vertices of the image indicating that this configuration state is made of a disordered arrangement of +1/-1 spin ice charges, corresponding to an Ice I state (52-53).

b) **Superconducting characterization**

This rich magnetic scenario can be exploited to control the dynamics of superconducting vortex lattice using different knobs, each one with different functionalities. Following the previous analysis, there are three different properties of the honeycomb Co array that can be used to control superconducting vortex motion in this superconducting/magnetic hybrid system. First, the array provides a structural basis to nucleate magnetic topological defects with fixed spatial density and hexagonal symmetry. Second, black/white magnetic charges (+1/-1 spin ice charges) provide attractive/repulsive magnetic pinning potentials for superconducting vortices depending on $H_z$ orientation. Third, local magnetic configuration at the intersections of connected Co bars defines the position of magnetic half-vortices at each cell of the honeycomb array and controls the asymmetry of the magnetic pinning potential. The first two properties of the honeycomb Co array allow knowing whether or not the vortices accomplish a regular distribution along the array. The third condition turns out the clue to obtain a robust and protected ratchet effect.

We begin analyzing how we can control the vortex lattice motion. The particular vortex density is obtained applying the required magnetic field perpendicular to the sample. At temperatures close to the superconducting critical temperature ($T_c$) the artificially induced periodic potential wells overcome the pinning potentials induced by the random distribution of defects in the sample (54). Therefore, the moving vortex lattice could interact with the periodic array of pinning centers. Jaque et al. (55) studied the interplay between the superconducting vortex lattice and arrays of periodic nanobars. They found plateaux in the dissipation for specific values of the magnetic fields applied perpendicular to the sample. These plateaux are related to the periodicity of the array. The magnetoresistance, with applied magnetic field perpendicular to the sample, in the honeycomb array hybrid is shown in Fig. 3 (for comparison the usual monotonously increasing magnetoresistance of a plain Nb film is plotted in Fig. 3(c)). We do not observe plateaux, we observe evenly spaced minima when the Co honeycomb array is at remanence after applying a saturating magnetic field along the magnetic easy axis (see Fig. 3(b)), i.e. with the honeycomb array in an ordered Ice II configuration. Resistance minima are observed with an average spacing $\mu_0 H_1$ = 4.0 mT (as shown in the inset of Fig. 3(b)). This finding corresponds to the matching between the vortex lattice and the vertices in the array, as sketched in Fig. 3(a). Therefore, the vertices in the array act as magnetic pinning potentials. Each time the density of superconducting vortices is an integer number of the density of magnetic pinning centers the superconducting vortex lattice motion slows down, a resistance minimum appears and dissipation decreases. These sharp minima are the footprint of matching effect between the vortex lattice and the triangular unit cell of the charged sublattice (56). Therefore, the ordered spin-ice charge array allows controlling the vortex lattice motion. For the fabricated Co honeycomb lattice, the distance (a) between alternating vertices in the triangular cell (i.e. between spin ice charges of the same sign) is a = 765 nm that corresponds to the first matching field $\mu_0 H_1$ = 1.156 $\Phi_0/a^2$ = 4.02 mT. Thus, the experimental matching field $\mu_0 H_1$ = 4.0 mT is in good agreement with the calculated matching conditions in the ordered spin ice II configuration. We have to point out that the interaction which governs this behavior is between magnetic stray fields in the honeycomb array vertices (+1 /-1 charges) and the superconducting vortices (57). The ordered Ice II state provides an effective magnetic pinning potential for the superconducting vortex lattice when it matches either the triangular lattice of -1 ice charges (downward magnetic applied fields) or the triangular lattice of +1 ice charges (upward magnetic applied fields). On the contrary, when the Co honeycomb array is in an Ice I configuration, equally spaced resistance minima disappear, as can be observed in the magnetoresistance curve (see Fig. 3(d)). That is, matching effects between spin ice charges and the superconducting vortex lattice fade away due to the loss of long range order in Ice I phase: the triangular lattice of superconducting vortices at the first matching field ($H_1$) is randomly attracted/repelled by the positive/negative magnetic charges at the intersections of the honeycomb lattice resulting in a negligible synchronized pinning effect. In summary, the superconducting vortex dynamics can be controlled using the magnetic history of the hybrid superconducting/magnetic sample.

**RATCHET EFFECTS AND DISCUSSION**

Once, we have identified the acting magnetic potentials, located in the honeycomb vertices with +1/-1 magnetic charges, we present the experimental behavior and the outcomes of our design ratchet device. As was quoted before, spatial asymmetries in the magnetic pinning potentials can be probed by superconducting vortex ratchet measurements (19, 25-27). First, we obtain the superconducting vortices applying perpendicular magnetic fields at matching conditions $H_z = H_1$. Next an ac current creates an alternating Lorentz force $F_L$ on the vortex lattice that results in a rectified vortex velocity, as long as there is an asymmetry between backward/forward pinning forces. In short, an ac current density $J = J_{ac} \sin(\omega t)$ is injected, where $\omega$ is the ac frequency, in our case 1 kHz and $t$ is time. This yields an alternating Lorentz force ($F_L$) on the vortices $\mathbf{F_L} = \mathbf{J} \times \Phi_0 \mathbf{z}$, $\Phi_0$ and $\mathbf{z}$ being the magnetic fluxoid and the unit vector parallel to the applied magnetic field respectively. Albeit the time averaged force on the vortices is zero, taking into account the Josephson expression (58) ($\mathbf{E} = \mathbf{B} \times \mathbf{v}$, being $\mathbf{E}$, $\mathbf{B}$ and $\mathbf{v}$ the electric field, the magnetic field and the vortex lattice velocity, respectively) an output dc voltage is measured proportional to the rectified vortex velocity. In summary, an ac current input yields a dc voltage output and a ratchet effect is achieved if forward/backward pinning forces are asymmetric. Fig. 4 shows the experimental results both when the honeycomb array is in an ordered Ice II state (Fig. 4(a)) and in a disordered Ice I state (Fig. 4(b)). In both cases, a clear positive ratchet voltage of several μV is observed, which is the characteristic outcome for interacting particles moving on asymmetric potentials. Thus, in spite of the very different magnetic configuration, our hybrid Co honeycomb/Nb device works in both cases as a typical rectifier device: input alternating forces generate output net flow. This is the most noteworthy finding of the present work: a ratchet effect is measured with the Co honeycomb array in a disordered Ice I state. Remarkably, long range order of the asymmetric pinning potentials is not necessary to obtain vortex velocity rectification.

To figure out the origin of this behavior we have to take into account the geometrical distribution of the magnetic half vortices comprising two Néel walls at each vertex of the honeycomb lattice, and this has to be done according to the ice rules. We can obtain a rough sampling of the half vortex geometrical distribution analyzing the MFM experimental data for Ice II and Ice I, ordered and disordered states, respectively. As indicated in Fig. 2(b), direct comparison between MFM experimental images and simulated MFM contrast allows establishing the average magnetization orientation at individual Co bars at each vertex in the honeycomb lattice. This procedure is carried on larger images taking into account ice-rules and half vortex asymmetries to draw the magnetization vectors in a consistent way both in ordered (Fig. 5) and disordered configurations (Fig. 6). Then, at each intersection, the orientation of magnetic half vortices is univocally determined by the local magnetic configuration, i.e. by the intersection edge at which the $-\pi$ rotation of the half vortex is localized. In brief, in the case of ordered Ice II configuration (Fig. 5), +1/-1 ice charges are arranged in a triangular lattice, existing only two kinds of magnetic half vortices in the image (see Fig. 5(d)) black and white. The V-shaped pairs of domain walls of these two half vortices point in opposite directions but, due to their opposite magnetic charges (+1 and -1), both of them provide magnetic potentials with the same asymmetry for vortices travelling along the easy axis, as shown in the simulated profile of Fig. 1(d). Thus, the ordered configuration of black/white half vortices in Ice II state is consistent with the net rectified vortex velocity observed in the experimental results of Fig. 4(a).

On the other hand, in the disordered Ice I configuration (Fig. 6), there is not long range order in the configuration of +1/-1 ice charges and different orientations of the magnetic half vortices can be observed. In particular, out of the 12 possible half vortex configurations within a honeycomb array (sketched in the insets of Fig. 6) we observe only 6 in the experimental image, 3 white and 3 black. This indicates that the experimental Ice I state is not in a fully random isotropic state but that it retains a certain global asymmetry, derived from its magnetic history. Experimental and simulated MFM profiles shown in Fig. 7 indicate that individual magnetic half vortices provide asymmetric pinning potentials for superconducting vortices travelling across those half vortices in any directions. Thus, we have obtained a robust and resilient ratchet device which works independently of the magnetic history of the device.

## SUMMARY

We have designed, fabricated and measured a superconducting ratchet device using, as the origin of the needed asymmetric potentials, magnetic half vortices with charged Néel walls in a spin ice honeycomb array and superconducting vortices driven by alternating forces as out-of-equilibrium particles. Magnetic half vortices are topologically confined at the honeycomb lattice intersections but their global configuration depends on spin ice states generated by magnetic frustration in the Co honeycomb arrays. The interplay among superconducting vortices, magnetic frustration, topology and spin ice states lead to a rich experimental scenario. Eventually our device can be controlled with three distinct topological defects, each one with a different functionality. We have superconducting vortices, +1 /-1  magnetic charges in the spin-ice with their associated stray fields, and  -1/2 half-magnetic vortices linked to a couple of charged Néel walls in each vertex of the Co honeycomb array. It is found that when superconducting vortices are pushed by zero average alternating forces, a net flow is always measured, independent of the magnetic history of the sample. Therefore a proof of concept of a robust and resilient interacting particles ratchet device has been developed. The mechanism responsible for the ratchet effect is independent of whether the sample is in an ordered (Ice II) or in a disordered state (Ice I). In both cases, the ratchet effect is generated by the asymmetry in the magnetic potential due to the asymmetric profile of the charged Néel walls that compose each magnetic half vortex.

## Acknowledgement

We thank support from Spanish MINECO grants FIS2016-76058 (AEI/FEDER, UE), Spanish CM grant S2013/MIT-2850. IMDEA Nanociencia acknowledges support from the 'Severo Ochoa' Programme for Centres of Excellence in R&D (MINECO, Grant SEV-2016-0686). DG acknowledges RYC-2012-09864, S2013/MIT-3007 and ESP2015-65597-C4-3-R for financial support and AG acknowledges financial support from Spanish MINECO Grant ESP2015-65597-C4-1-R.

**References**

1. Simon S M, Peskin C S and Oster G F. *What drives the translocation of proteins?.* **Proc Natl Acad Sci U S A. 89**, 3770 (1992).
2. Bar-Nahum G, Epshtein V, Ruckenstein AE, Rafikov R, Mustaev A , Nudler E. *A ratchet mechanism of transcription elongation and its control.* **Cell 120,** 183 (2005)
3. Rousselet J, Salome l, Ajdari A, Prost J. *Directional motion of brownian particles induced by a periodic asymmetric potential.* **Nature 370,** 446 (1994)
4. Marquet C, Buguin A, Talini L, Silberzan, P. *Rectified motion of colloids in asymmetrically structured channels.* **Phys. Rev. Lett. 88**, 168301 (2002)
5. Jung P, Kissner J G, Hänggi P. *Regular and Chaotic Transport in Asymmetric Periodic Potentials: Inertia Ratchets* **Phys. Rev. Lett. 76**, 3436 (1996)
6. Faucheux L P, Bourdieu L S, Kaplan, P. D. and Libchaber, A. J. *Optical thermal ratchet.* **Phys. Rev. Lett. 74**, 1504 (1995).
7. van Oudenaarden A and Boxer S G. *Brownian ratchet: molecular separations in liquid bilayers supported on patterned arrays*. **Science 285**, 1046 (1999)
8. Linke, H., Humphrey T. E. , Löfgren A., Sushkov, A. O., Newbury, R. Taylor R. P., Omling P**.** *Experimental tunneling ratchets.* **Science 286**, 2314 (1999)
9. Matthias S and Muller F. *Asymmetric pores in a silicon membrane acting as massively parallel brownian ratchets.* **Nature 424**, 53 (2003)
10. Costache M V, Valenzuela S O. *Experimental spin ratchet*. **Science 330,** 1645 (2010)
11. Gliga S, Hrkac, G., Donnelly C, Büchi J , Kleibert A, Cui J, Farhan A, Kirk E, Chopdekar R V , Masaki Y, Bingham N S, Scholl A, Stamps R L and Heyderman L J. *Emergent dynamic chirality in a thermally driven artificial spin ratchet*. **Nat. Mat. 16**, 1106 (2017)
12. Perez-Junquera A, Marconi V I, Kolton A B, Alvarez-Prado, L M, Souche Y, Alija A, Velez M, Anguita J V, Alameda J M, Martín J I and Parrondo J M R. *Crossed-ratchet effects for magnetic domain wall motion* **Phys. Rev. Lett. 100**, 037203 (2008)
13. Auge, A. Weddemann, A. Wittbracht, F. Hutten, A. *Magnetic ratchet for biotechnological applications***. Appl. Phys. Lett. 94**, 183507 (2009)
14. Franken J H, Swagten H J M, Koopmans B. *Shift registers based on magnetic domain wall ratchets with perpendicular anisotropy.* **Nat. Nanotech. 7**, 499, (2012)
15. Lavrijsen R, Lee J H, Fernandez-Pacheco A, Petit D C M C, Mansell R and Cowburn, R.P. *Magnetic ratchet for three-dimensional spintronic memory and logic* **Nature 493**, 647 (2013).
16. Mochizuki M, Yu X Z, Seki S, Kanazawa N, Koshibae W, Zang J, Mostovoy M, Tokura Y and Nagaosa, N. *Thermally driven ratchet motion of a skyrmion microcrystal and topological magnon Hall effect.* **Nat. Mat. 13**, 241 (2014)
17. Zapata I, Bartussek R, Sols F, Hanggi P. *Voltage rectification by a SQUID ratchet* **Phys. Rev. Lett. 77**, 2292 (1996).
18. Lee C S, Janko B, Derenyi I, Barabasi A L. *Reducing vortex density in superconductors using the 'ratchet effect'* .**Nature 400**, 337 (1999)
19. Villegas J E, Savelev S, Nori F, Gonzalez E M, Anguita J V, Garcia R and J. L. Vicent J L. *A Superconducting Reversible Rectifier that Controls the Magnetic Flux Quanta*

**Science 302**, 1188 (2003)

20. Beck M, Goldobin E, Neuhaus M, Siegel M, Kleiner R and Koelle D. *High-efficiency deterministic Josephson vortex ratchet.* **Phys. Rev. Lett. 95**, 090603 (2005)
21. Van de Vondel J, Silva C C D, Zhu B Y, Morelle M and Moshchalkov V V. *Vortex-rectification effects in films with periodic asymmetric pinning* **Phys. Rev. Lett. 94**, 057003 (2005)
22. Togawa Y, Harada K, Akashi T, Kasai H, Matsuda T, Nori F, Maeda A, and Tonomura A. *Direct Observation of Rectified Motion of Vortices in a Niobium Superconductor* **Phys. Rev. Lett. 95**, 087002 (2005).
23. Wambaugh J F, Reichhardt C, Olson C J, Marchesoni F and Nori F. *Superconducting fluxon pumps and lenses* **Phys. Rev. Lett. 83**, 5106 (1999).
24. Ji J D, Yuan J, He G, Jin B A B, Zhu B Y, Kong X D, Jia X Q, Kang L, Jin K, and Wu P. *Vortex ratchet effects in a superconducting asymmetric ring-shaped device* **Appl. Phys. Lett. 109**, 242601 (2016).
25. de Souza Silva CC, Silhanek AV, Van de Vondel J, Gillijns W, Metlushko V, Ilic B, Moshchalkov V. *Dipole-induced vortex ratchets in superconducting films with arrays of micromagnets.* **Phys Rev Lett. 98**, 117005 (2007)
26. Perez de Lara D, Castaño F J, Ng B G, Korner H S, Dumas R K, Gonzalez E M, Liu K, Ross C A, Schuller I K and J. L. Vicent. *Rocking ratchet induced by pure magnetic potentials with broken reflection symmetry* **Phys. Rev. B 80,** 224510 (2009).
27. Gomez A, Gonzalez E M, Iglesias M, Sanchez N, Palomares F J, Cebollada F, Gonzalez J M and J L Vicent J L**.** *A superconducting/magnetic hybrid rectifier based on Fe single-crystal nanocentres: role of magnetic and geometric asymmetries.***J. Phys. D: Appl. Phys. 46**, 095302 (2013).
28. Nisoli C, Moessner R and Schiffer P. *Colloquium: Artificial spin ice: Designing and imaging magnetic frustration* **Rev. Mod. Phys. 85**, 1473 (2013)
29. Castelnovo C, Moessner R and Sondhi S L. *Magnetic monopoles in spin ice.* **Nature 451**, 42 (2008)
30. Latimer M L, Berdiyorov G R, Xiao Z L, Peeters F and Kwok, W. K. *Realization of artificial ice systems for magnetic vortices in a superconducting MoGe thin film with patterned nanostructures* **Phys. Rev. Lett. 111***,* 067001 (2013).
31. Trastoy J, Malnou M, Ulysse C, Bernard R, Bergeal N, Faini G, Lesueur J, Briatico J and Villegas J E. *Freezing and thawing of artificial ice by thermal switching of geometric frustration in magnetic flux lattices.* **Nat. Nanotech. 9**, 710 (2014).
32. Ge J Y, Gladilin V N, Tempere J, Zharinov V S, Van de Vondel J, Devreese J T, and Moshchalkov V V. *Direct visualization of vortex ice in a nanostructured superconductor* **Phys. Rev. B 96,** 134515 (2017).
33. Wang Y L, Ma X, Xu J, Xiao Z L, Snezhko A, Divan R, Ocola L E, J. Pearson J E, Janko B and Kwok W K. *Switchable geometric frustration in an artificial spin-ice–superconductor hetero system* **Nat. Nanotech. 13**, 10.1038/s41565-018-0162-7 (2018).
34. Tchernyshyov O and Chern G –W. *Fractional Vortices and Composite Domain Walls in Flat Nanomagnets*, **Phys. Rev. Lett. 95**, 197204 (2005)

35. Rodríguez-Rodríguez G, Rubio H, Vélez M, Pérez-Junquera A, Anguita J V, Martín J I and Alameda J M. *Closure magnetization configuration around a single hole in a magnetic film,* **Phys. Rev. B 78**, 174417 (2008).
36. del Valle J, Gomez A, Gonzalez E M, Osorio M R, Granados D and J L Vicent. *Superconducting/magnetic three-state nanodevice for memory and reading applications* **Sci. Rep. 5**, 15210 (2015).
37. Vansteenkiste A, Leliaert J, Dvornik M, Helsen M, Garcia-Sanchez F and Van Waeyenberge B. *The design and verification of MuMax3.* **AIP Advances 4**, 107133 (2014).
38. Rouco V, Córdoba R, De Teresa J M, Rodríguez L A, Navau C, Del-Valle N, Via G, Sánchez A, Monton C, Kronast F, Obradors X, Puig T and Palau A. *Competition between Superconductor – Ferromagnetic stray magnetic fields in $YBa_2Cu_3O_{7-x}$ films pierced with Co nano-rods* **Sci. Rep. 7**, 5663 (2017).
39. G. E. Rowland, MuView, http://www.grahamerowlands.com/main/muview
40. Hierro-Rodriguez A, Cid R, Vélez M, Rodriguez-Rodriguez G, Martín J I, Alvarez- Prado L M, Alameda M. *Topological Defects and Misfit Strain in Magnetic Stripe Domains of Lateral Multilayers With Perpendicular Magnetic Anisotropy.* **Phys. Rev. Lett. 109**, 117202 (2012).
41. Loehr J, Loenne M, Ernst A, de las Heras D and Fischer T M. *Topological protection of multiparticle dissipative transport*. **Nat. Comm. 7**, 11745 (2016).
42. Harris M J, Bramwell S T, McMorrow D F, Zeiske T and Godfrey K W. *Geometrical frustration in the ferromagnetic pyrochlore $Ho_2Ti_2O_7$* **Phys. Rev. Lett. 79**, 2554 (1997).
43. Wang R F, Nisoli C, Freitas R S, Li J, McConville W, Cooley B J, Lund M S, Samarth N, Leighton C, Crespi V H and Schiffer P. *Artificial 'spin ice' in a geometrically frustrated lattice of nanoscale ferromagnetic islands* **Nature 439,** 303 (2006)
44. Tanaka M, Saitoh E, Miyajima H, Yamaoka T and Iye Y. *Magnetic interactions in a ferromagnetic honeycomb nanoscale network*. **Phys. Rev. B 73**, 052411 (2006).
45. Möller G and Moessner R. *Magnetic multipole analysis of kagome and artificial spin-ice dipolar arrays* **Phys. Rev. B 80**, 140409(R) (2009).
46. Qi Y, Brintlinger T and Cumings J. *Direct observation of the ice rule in an artificial kagome spin ice* **Phys. Rev. B 77**, 094418 (2008)-
47. Bhat V S, Sklenar J, Farmer B, Woods J, Hastings J T, Lee S J, Ketterson J B and De Long L E. *Controlled Magnetic Reversal in Permalloy Films Patterned into Artificial Quasicrystals* **Phys. Rev. Lett. 111**, 077201 (2013)
48. Zeissler K, Chadha M, Lovell E, CohenL F and Branford W R**.** *Low temperature and high field regimes of connected kagome artificial spin ice: the role of domain wall topology.* **Sci. Rep**. **6**, 30218 (2016).
49. Sun L, Zhou C, Liang J H, Xing T, Lei N, Murray P, Liu K, Won C, Wu Y Z. *Magnetization reversal in kagome artificial spin ice studied by first-order reversal curves*. **Phys. Rev. B 96**, 144409 (2017).
50. Pushp A, Phung T, Rettner C, Hughes B P, Yang S H, Thomas L and Parkin S P. *Domain wall trajectory determined by its fractional topological edge defects*. **Nat. Phys. 9** 505 (2013).
51. Huber A and Schäfer R. *Magnetic Domains* (Springer-Verlag, Berlin Heidelberg 1998).

52. Möller G and R. Moessner R. *Magnetic multipole analysis of kagome and artificial spin-ice dipolar arrays.* **Phys. Rev. B 80** 140409(R) (2009).
53. Mellado P, Petrova O, Shen Y and O. Tchernyshyov O**.** *Dynamics of Magnetic Charges in Artificial Spin Ice.* **Phys. Rev. Lett. 105** 187206 (2010).
54. Velez M, Jaque D, Martín J I, Guinea F and Vicent J L. *Order in Driven Vortex Lattices in Superconducting Nb Films with Nanostructured Pinning Potentials*.**Phys. Rev. B 65**, 094509 (2002).
55. Jaque D, Gonzalez E M, Martín J I, Anguita J V and J. L. Vicent J L. *Anisotropic Pinning Enhancement in Nb Films with Arrays of Submicrometric Ni Lines* **Appl. Phys. Lett. 81**, 2851 (2002).
56. Vélez M, Martín J I, Villegas J E, Hoffmann A, González E M, Vicent J L and I. K. Schuller**.** *Periodic Pinning Effects in Superconducting Films with Ordered Arrays of Magnetic Dots* **J. Magn. Magn. Mat. 320**, 2547 (2008) and references therein.
57. Gomez A, Gilbert D A, Gonzalez E M, Liu K and J. L. Vicent J L. *Control of dissipation in superconducting films by magnetic stray fields.* **Appl. Phys. Lett. 102,** 052601 (2013)
58. Josephson B D. *Potential differences in mixed state of type 2 superconductors.* **Phys. Lett. 16**, 242 (1965).

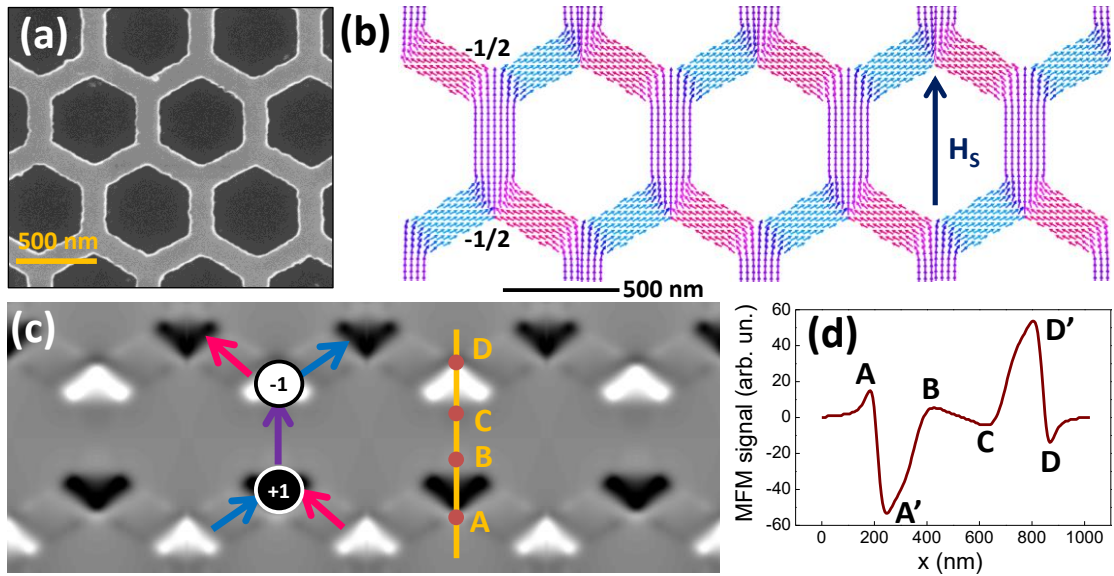

*Figure 1:* **Micromagnetic configuration of the honeycomb array**. (a) SEM image of Co honeycomb array. (b) Micromagnetic simulation of Co honeycomb array at easy axis remanence. Note the presence of -1/2 half vortices at opposite bar sides. (c) Simulated MFM contrast image from the micromagnetic configuration in (b) at 50 nm lift height. Sketch shows average magnetization direction at each bar and spin ice charge at the intersection. (d) Contrast profile along the vertical line marked in (c).

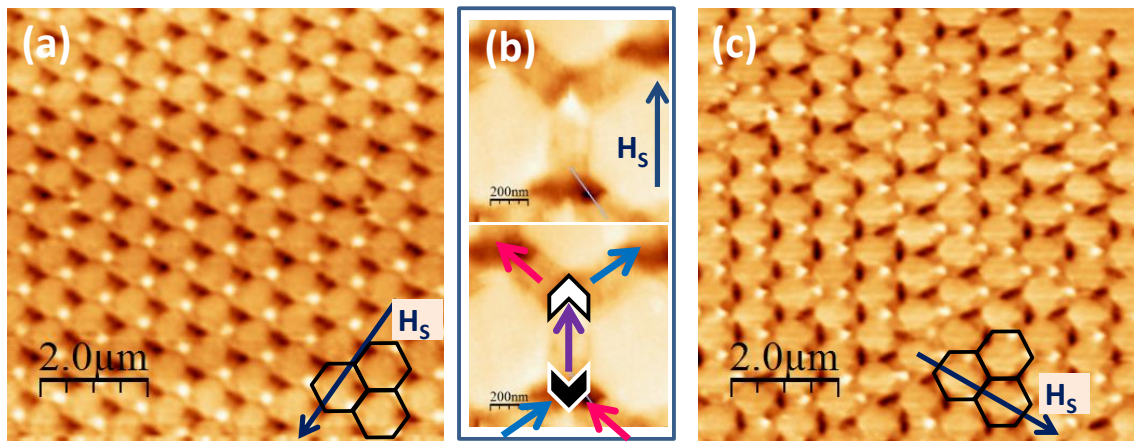

*Figure 2:* **MFM images of the honeycomb array at different remanent states**. (a) Easy axis remanence. Note the ordered arrangement of white/black spin ice charges corresponding to an Ice II state. (b) Detail of remanent configuration of a single bar in the array. Note the V shaped pairs of Neel walls that meet at each bar end corresponding to magnetic half vortices. The lower part of the image shows a sketch of magnetization configuration in the single bar: arrows indicate magnetization direction, V shapes represent the pair of Neel walls with the half vortex core on the tip and black/white color depending on the sign of the ice charge at the intersection (+/- 1). (c) Hard axis remanence made up of a disordered mixture of white/black spin ice charges of similar intensity corresponding to an Ice I state.

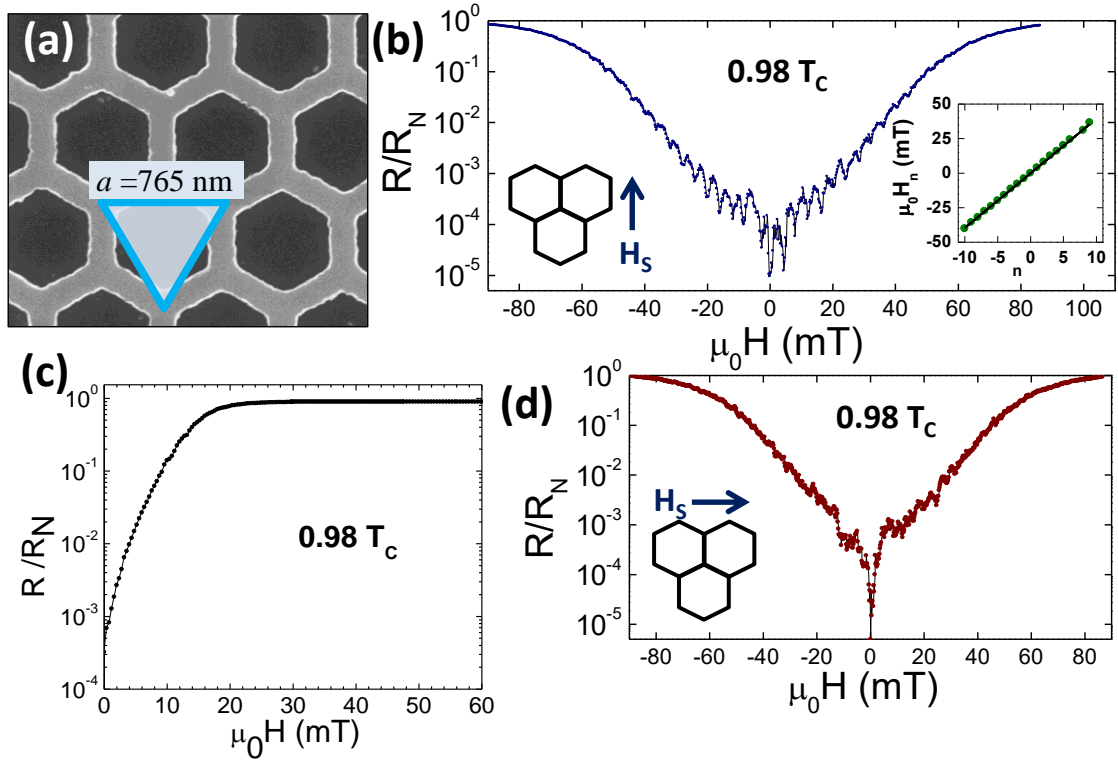

*Figure 3:* **Superconducting vortices dynamics as a function of order/disorder in the spin ice system.** (a) Scanning Electron Microscopy image of Co honeycomb array with triangle indicating the geometrical dimensions of the lattice of -1 ice charges in Ice II state. (b) Normalized magnetoresistance curve of the hybrid device at $0.98T_C$ after saturating the Co honeycomb array with $H_S$ along the magnetic easy axis (ordered Ice II configuration). Note the periodic minima in the resistance at regular field intervals $\mu_0H_n$. Inset shows $\mu_0H_n$ *vs.* n linear dependence with slope 4 mT. (c) Normalized magnetoresistance curve of a plain Nb film at $0.98T_C$. (d) Normalized magnetoresistance curve of the hybrid device at $0.98T_C$ after saturating the Co honeycomb array with $H_S$ perpendicular to the magnetic easy axis (disordered Ice I configuration). Note the absence of regular magnetoresistance minima in contrast with the behavior observed in (b).

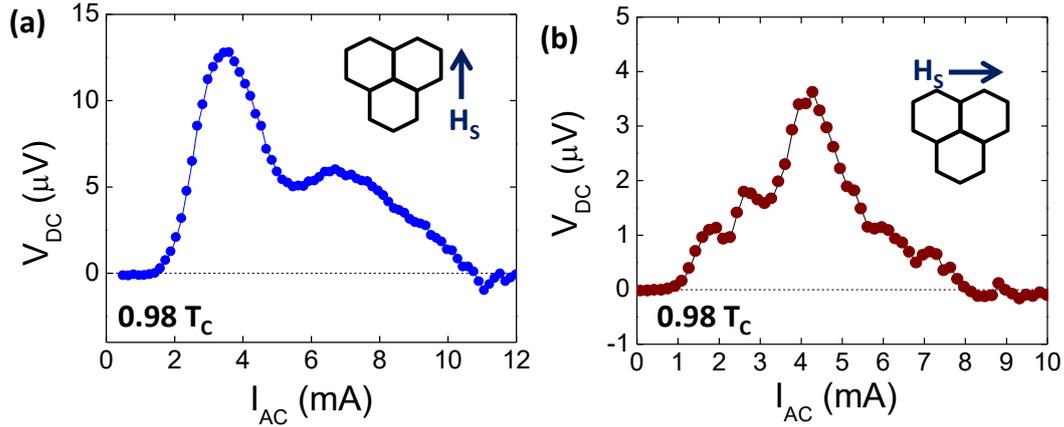

*Figure 4:* **Rectification of superconducting vortex motion by Co honeycomb array.** Rectified ratchet voltage in the hybrid device at $H_z = H_1$ after two different saturation field configurations: (a) $\mu_0 H_S = 7$ T parallel to easy axis and (ordered Ice II state) (b) $\mu_0 H_S = 7$ T perpendicular to the magnetic easy axis (disordered Ice I configuration).

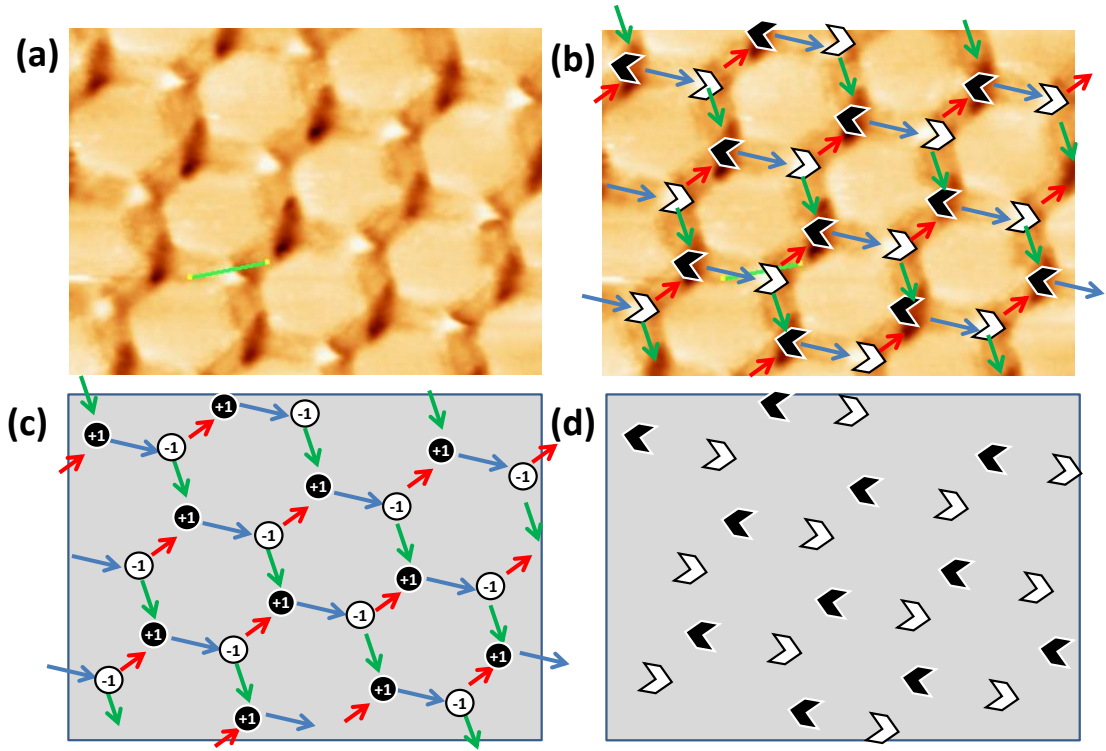

*Figure 5:* **Analysis of topological defects from MFM image in ordered Ice II configuration.** (a) MFM image of honeycomb array. (b) Sketch of local magnetization orientation and half vortex position. (c) Sketch of magnetization configuration and spin ice charges. (d) Sketch of configuration of magnetic half vortices. Note that in this ordered Ice II configuration +1 (or -1) ice charges are arranged in a hexagonal lattice and there are only two kinds of magnetic half vortices in the image.

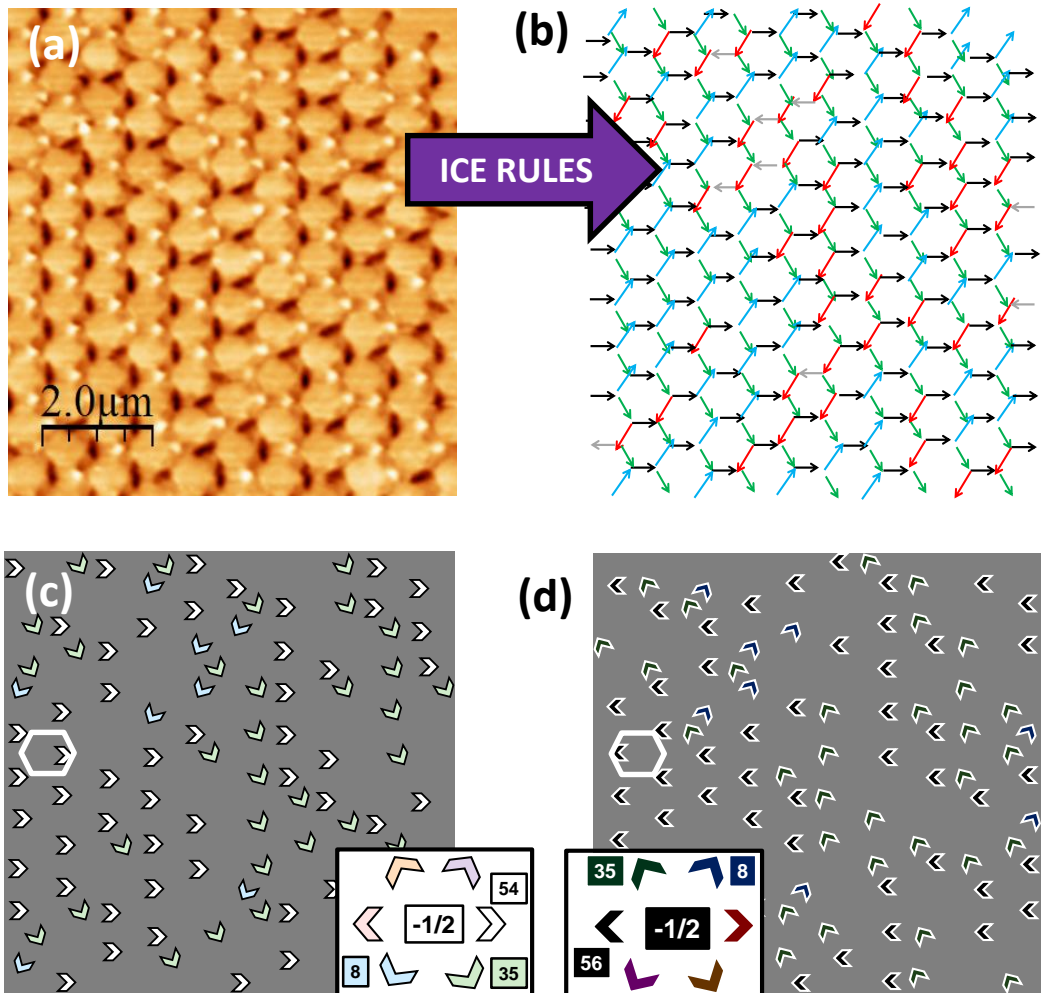

*Figure 6:* **Analysis of topological defects from MFM image in disordered Ice I configuration.** (a) MFM image of honeycomb array (see Fig. 2(c)). (b) Sketch of local magnetization orientation derived from (a) using ice rules. Sketches of position and orientation of white (c) and black (d) magnetic half vortices derived from (a). Insets show a sketch of all the possible half vortex configurations in a honeycomb lattice and numbers in squares indicate the actual count for each kind of half vortex present in (c-d). Note the absence of long range order in (c) and (d). Interestingly, only 3 out of 6 orientations are present. The contribution of the same orientation, but different (+1/-1) magnetic charges adds in the ratchet effect.

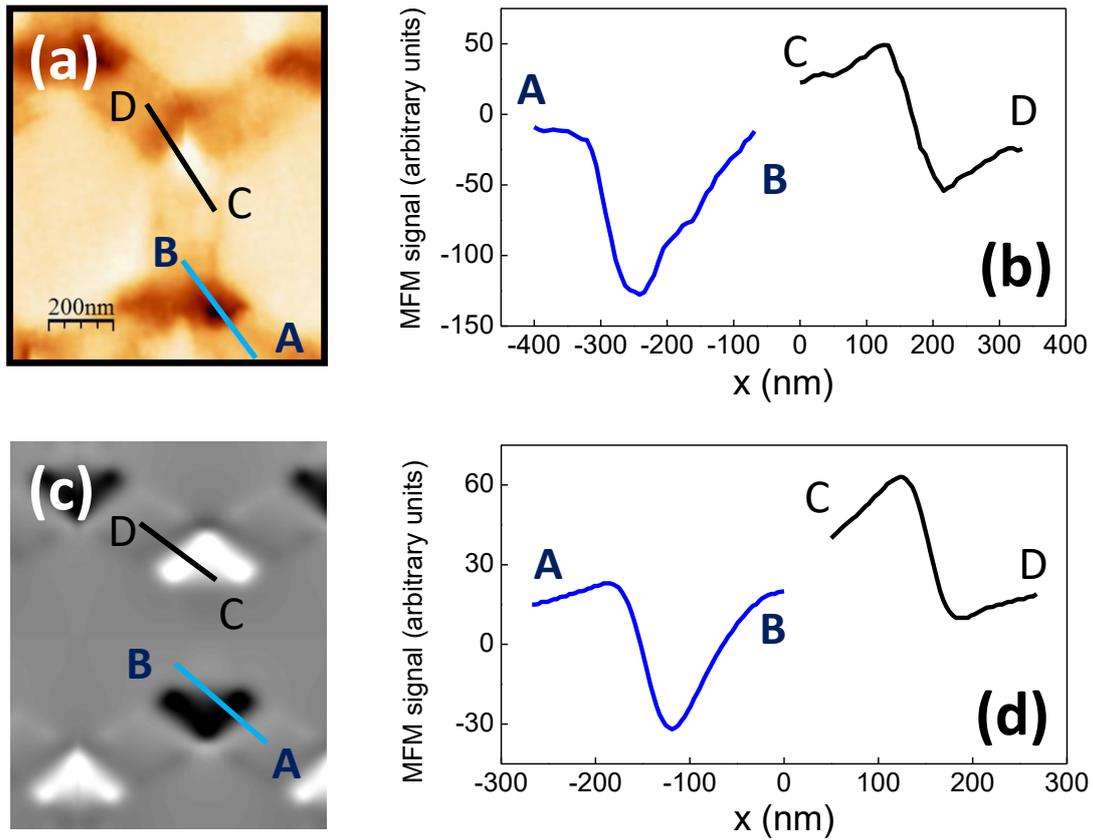

*Figure 7:* **Experimental and simulated potential profiles.** (a) Experimental MFM image of a single bar in the array. (b) AB and CD profiles from experimental MFM image in (a). (c) Simulated MFM image of a single bar in the array. (d) AB and CD profiles from simulated MFM image in (c). Note the clear asymmetry upon crossing the Neel walls that emerge from -1/2 edge vortices (steep descending vs. gradual ascending slopes) with the same sign in AB and CD profiles.